\documentclass[conference]{IEEEtran}
\IEEEoverridecommandlockouts

\usepackage{cite}
\usepackage{amsmath,amssymb,amsfonts}
\usepackage{algorithmic}
\usepackage{graphicx}
\usepackage{textcomp}
\usepackage{xcolor}
\usepackage{booktabs}
\usepackage{url}
\def\BibTeX{{\rm B\kern-.05em{\sc i\kern-.025em b}\kern-.08em
    T\kern-.1667em\lower.7ex\hbox{E}\kern-.125emX}}
\begin{document}

\title{Micro-Segmentation Anomaly Detection in Zero-Trust Software-Defined Network Fabrics\\
}

\author{
\IEEEauthorblockN{Ashly Joseph}
\IEEEauthorblockA{
IEEE Member\\
ashlyjoseph.me@gmail.com
}
\thanks{
Author-accepted manuscript. Published as:
A. Joseph, ``Micro-Segmentation Anomaly Detection in Zero-Trust
Software-Defined Network Fabrics,'' in \emph{2025 Artificial
Intelligence and Smart Technologies for Sustainability Conference
(AISTS)}, Rajkot, India, 2025,
doi: 10.1109/AISTS66100.2025.11233161.

\par\medskip

\textcopyright{} 2025 IEEE. Personal use of this material is permitted.
Permission from IEEE must be obtained for all other uses, in any current
or future media, including reprinting/republishing this material for
advertising or promotional purposes, creating new collective works, for
resale or redistribution to servers or lists, or reuse of any copyrighted
component of this work in other works.
}
}

\maketitle
\begin{abstract}

Zero Trust Architecture (ZTA) principles need rigorous network segmentation and ongoing verification to reduce implicit trust and lateral threat propagation. This paper investigates anomaly detection in software-defined networking (SDN) systems by micro-segmentation, using deep learning models to detect harmful actions that evade traditional coarse-grained monitoring. Two models are developed: a Vision Transformer (ViT) and a 1D Convolutional Neural Network (1D-CNN), which are used to both raw and micro-segmented network flow data. Experimental findings from a simulated zero-trust SDN dataset indicate that micro-segmentation substantially improves detection accuracy. The models trained on segmented input demonstrate enhanced accuracy and F1-scores (F1 = 0.95) compared to those utilizing unsegmented raw data (F1 = 0.90). The ViT-based detector marginally surpasses the 1D-CNN, particularly in recognizing nuanced lateral movement patterns that are unnoticed in unprocessed data. These findings highlight the significance of including micro-segmentation inside zero-trust networks to enhance intrusion detection efficacy. Future efforts will broaden this methodology to include extensive real-world network datasets and dynamic online segmentation techniques.
\end{abstract}


\section{Introduction}

Businesses are progressively embracing Zero Trust cybersecurity concepts, which operate on the premise of no implicit trust for any person or device and necessitate ongoing identity verification and access management.~\cite{zeadally2022network} A core element of Zero Trust Architecture (ZTA) is micro-segmentation, which involves partitioning a network into detailed, secure segments or "micro-perimeters" to restrict lateral movement by attackers.~\cite{Ni2022}~\cite{Kang2022} By isolating resources and enforcing access controls inside each segment, any breach is limited, so preventing hackers from navigating the wider network.~\cite{Sheikh2021} This method substantially diminishes the attack surface and aids in containing possible breaches. 

Nonetheless, segmentation alone is inadequate when advanced threats circumvent basic protections. Conventional micro-segmentation methods impose structural limitations on communication but frequently lack sophisticated anomaly detection features.~\cite{Basta2022} Adversaries utilizing covert or resource-exploitation methods may function inside authorized segment limits, circumventing signature-based detection systems. This necessitates the implementation of advanced anomaly detection technologies integrated inside each micro-segment, consistent with the Zero Trust principle of "never trust, always verify." Intrusion Detection Systems (IDS) that continually monitor intra- and inter-segment traffic can identify anomalous patterns in real time.~\cite{Gamage2020}

The advent of Software-Defined Networking (SDN) enhances this concept by providing centralized management and oversight of network traffic, but introducing additional dangers, including possible vulnerabilities in the control plane.~\cite{Ding2018} Incorporating anomaly detection into SDN-based zero-trust frameworks is crucial for the prompt identification and alleviation of security concerns.~\cite{Yang2023} In this regard, deep learning techniques have demonstrated efficacy in Network Intrusion Detection Systems (NIDS) owing to their capacity to simulate intricate behavioral patterns.~\cite{Dosovitskiy2020}

Convolutional Neural Networks (CNNs) have demonstrated success in autonomously extracting traffic characteristics and surpassing conventional classifiers in the detection of malware and intrusions.~\cite{Vaswani2017} Furthermore, attention-based architectures, particularly Transformers and Vision Transformers (ViTs), have attained exceptional performance in sequence modeling tasks. Vision Transformers (ViTs) have been modified for network security by representing network flows as sequences or pictures, resulting in competitive performance.~\cite{DeRose2024}
This paper provides an in-depth analysis of the significance of micro-segmentation in improving anomaly detection inside zero-trust, SDN-based network settings. Two deep learning-based network intrusion detection system models a Vision Transformer (ViT) and a one-dimensional convolutional neural network (1D-CNN) are constructed and assessed using both raw and micro-segmented network traffic data. Their performance is evaluated to measure the effect of micro-segmentation on anomaly detection skills. Traditional classifiers have difficulties in identifying minor lateral movements within flat network topologies, but attention-based models such as ViT and CNNs can utilize segment-specific patterns to enhance detection precision. This study examines if the inclusion of micro-segmentation context may significantly improve the sensitivity of anomaly detection.

This study's contributions are as follows:
\begin{itemize}
    \item A simulated zero-trust Software-Defined Networking environment is established, producing labeled traffic data that includes insider risks and lateral movement scenarios.
    \item The ViT and 1D-CNN model architectures are engineered for anomaly detection in time-series network flow data.
    \item It is shown that micro-segmentation markedly enhances performance measures, including accuracy, F1-score, and AUC. Numerous qualitative instances are presented in which micro-segmented models identify hazards that are neglected in raw data.
\end{itemize}

\section{Related Work}

\subsection{Zero Trust and Micro-Segmentation Security}
Zero Trust Architecture (ZTA) has been defined in NIST SP 800-207~\cite{Rose2019}, highlighting continuous verification and micro-segmentation as critical strategies for protecting network assets.~\cite{Ahmed2025} Ni et al. classify micro-segmentation methodologies into VM-based, network-infrastructure-based, and host-agent-based segmentation.~\cite{Ni2022} Initial host-agent methods offered robust isolation at the workload level but resulted in significant resource overhead on endpoints. Kang et al. proposed a distributed, IDS-driven micro-segmentation paradigm that dynamically generates fine-grained segments in response to detected abnormalities, therefore alleviating endpoint strain.~\cite{Kang2022} Sheikh et al. showed network-centric micro-segmentation via Illumio ASP, implementing whitelist policies among application layers.~\cite{Sheikh2021} Basta et al. conducted a quantitative comparison of micro-segmented and flat designs, demonstrating that segmentation significantly lowers attack vulnerability and enhances security metrics.~\cite{Basta2022}

\subsection{Anomaly Detection in SDN and Networks}
Software-Defined Networking (SDN) provides centralized management and extensive visibility, enabling robust telemetry acquisition for machine learning-driven intrusion detection.~\cite{Lashkari2016} Early Network Intrusion Detection Systems utilized support vector machines and clustering; however, deep learning methodologies have exhibited enhanced accuracy. Gamage and Samarabandu examine CNNs, RNNs (including LSTMs), and autoencoders for Network Intrusion Detection Systems (NIDS). Javaid et al. presented a hybrid CNN–LSTM model that captures spatial and temporal traffic characteristics, resulting in enhanced detection rates.\cite{Gamage2020} Attention-based models have been modified: Yang et al. created a Vision Transformer for NSL-KDD intrusion detection, whilst De Rose et al. introduced VINCENT, utilizing knowledge distillation to enhance ViT efficiency.~\cite{Yang2023} FlowTransformer exhibits pure Transformer models for flow-based anomaly detection. Notwithstanding these advancements, elevated false-positive rates and restricted generalization to unfamiliar assaults continue to exist. The integration of contextual information, including micro-segmentation, has demonstrated an improvement in detection specificity. However, these studies frequently depend on aggregated flow data without segment-awareness, so constraining their capacity to identify specific hazards. To tackle this issue, we compare our deep models with traditional machine learning techniques (e.g., Random Forest[7], SVM[15]), emphasizing the benefits derived from segmentation.

\section{Background and Problem Definition}
Zero-trust network architectures need ongoing authentication and permission for each access request.~\cite{Rose2019} Such fabrics are generally implemented by Software-Defined Networking (SDN), facilitating the dynamic implementation of least-privilege regulations. Micro-segmentation partitions the network into distinct zones ("micro-segments") based on workload function, data sensitivity, or user role, with inter-segment connectivity regulated by Policy Engine/Enforcement Points.~\cite{Ni2022}

The architecture provides improved visibility: all intra- and inter-segment traffic can be monitored and analyzed for irregularities. The detailed scope facilitates the collection of typical behavior profiles for each section, rendering deviations more apparent. Let F = {fi} represent the collection of observed network flows. In a raw-data method, flows are examined collectively without consideration of segment affiliation. I utilize a unified model that analyzes each flow vector accompanied by a one-hot segment identification.  This method allows the model to simultaneously learn global traffic patterns and segment-specific behaviors without the need for separate models for each segment.  In the ViT, the segment ID is included as an extra token; in the 1D-CNN, it enhances the input feature vector directly.

An anomaly detector \(\mathcal{D}_{\mathrm{seg}}\) that utilizing micro-segmentation context is hypothesized to outperform \(\mathcal{D}_{\mathrm{raw}}\), which functions on raw flows, especially for localized lateral-movement or low-and-slow attacks. By focusing on detailed segments, \(\mathcal{D}_{\mathrm{seg}}\) can learn each segment’s normal behavior and identify subtle abnormalities pertinent to that context.

Micro-segmentation facilitates dynamic policy responses upon anomaly detection. If \(\mathcal{D}_{\mathrm{seg}}\) flags a segment as compromised, the system can autonomously isolate that segment quarantining it by severing cross-segment communication as a containment strategy. Primary issues include managing increased data granularity (data scarcity per segment or increasing dimensionality) and sustaining low false-positive rates to prevent unnecessary isolations.

\begin{figure}[!t]
  \centering
  \includegraphics[width=\linewidth]{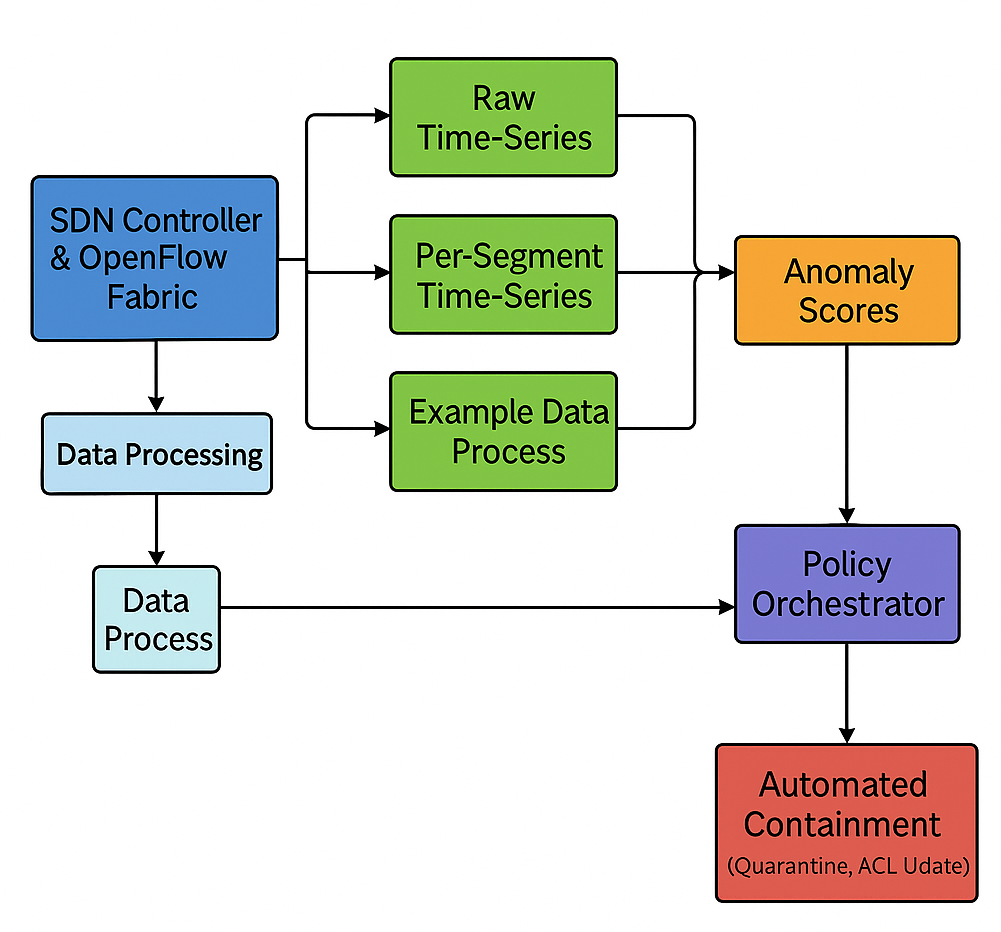}
  \caption{System architecture for micro-segmentation anomaly detection.}
  \label{fig:architecture}
\end{figure}

Figure 1 provides a comprehensive overview of the proposed anomaly detection pathway. Traffic telemetry is initially gathered from the SDN controller and forwarding fabric, thereafter preprocessed into three concurrent streams: raw network time-series, per-segment time-series, and an illustrated sample data process. Raw and micro-segmented data streams were individually evaluated utilizing 1D-CNN and ViT architectures to provide a cross-model comparison of segmentation strategies. The scores are integrated and evaluated against adaptive thresholds, with any alarms prompting the policy orchestrator to implement automatic containment measures, such segment quarantine or ACL modifications.

\section{Data and Pre-Processing}

\subsection{Dataset Simulation and Features}
To evaluate the anomaly detection framework, network traffic is simulated using real-world benchmarks, including CIC-IDS2017~\cite{CICIDS2017} and NSL-KDD.~\cite{Tavallaee2009} The simulated Software-Defined Network (SDN) architecture consists of several logical micro-segments (e.g., Web Server, Database, User Devices), each demonstrating characteristic intra-segment traffic patterns. Attack scenarios, such as port scans, lateral movement, data exfiltration, and denial-of-service, are incorporated such that each flow record includes a source/destination segment identification in addition to regular features.

Raw packet captures are consolidated into flow-level data at predetermined intervals (e.g., one minute). For each segment or host, vectors of byte and packet counts, unique connection counts, protocol distributions, and more metrics are calculated.  The segmented dataset preserves these attributes for each segment pair (both intra- and inter-segment), resulting in a feature matrix of dimensions $(T \times S)\times d$, where $T$ represents the number of time windows, $S$ denotes the number of segments, and $d$ indicates the feature dimension.  The raw dataset consolidates all segments into a single $T \times d$ matrix.   Both datasets undergo the same preprocessing procedures min-max scaling and, where applicable, categorical encoding along with class balancing using undersampling or SMOTE to reduce bias.

\subsection{Train–Test Splits and Labeling}
Data are temporally divided into training, validation, and test sets to simulate actual deployment: initial traffic provides the training set, whilst subsequent traffic with fresh assaults comprises the test set. Each time-window instance is categorized as \emph{normal} or \emph{anomalous}.  In the segmented view, anomalies indicate either intra-segment deviations (e.g., internal traffic surges) or inter-segment regulation breaches. The raw dataset uses a singular combined label, while the segmented dataset trains distinct instances for every segment. Subtle "low-and-slow" exfiltration instances, undetectable in the aggregated perspective, become detectable in the segmented context, offering a thorough assessment of micro-segmentation's advantages.

Each micro-segment in the SDN simulation has been defined using distinct IP subnets in Mininet, corresponding to logical responsibilities (e.g., Web Server: \texttt{10.0.1.0/24}, Database: \texttt{10.0.2.0/24}, User Devices: \texttt{10.0.3.0/24}). Segment labels were allocated during feature extraction according to source and destination IP addresses. The segment identifiers were one-hot encoded and added to each feature vector. In the ViT model, the segment ID was included as an extra embedding token, enabling the model to acquire attention patterns pertinent to segment context. A cohesive model architecture was used across segments instead of training an individual model for each section.

\subsection{Simulated Dataset Overview}
The simulation environment replicates a medium-scale corporate Software-Defined Networking (SDN) with five unique parts. We utilized CIC-IDS2017~\cite{CICIDS2017} and NSL-KDD~\cite{Tavallaee2009} attack profiles to produce traffic across a two-day simulated period. Essential statistics are included in Table~\ref{tab:dataset-summary}.

\begin{table}[h]
\caption{Simulated Dataset Overview}
\label{tab:dataset-summary}
\centering
\begin{tabular}{|l|l|}
\hline
\textbf{Parameter} & \textbf{Value} \\
\hline
Total flows & 250,000 \\
Segments & 5 (Web, DB, Auth, Users, DNS) \\
Time duration & 48 hours \\
Time window granularity & 1 minute \\
Benign/Attack ratio & 85/15 \\
Attack types & Lateral movement, Exfiltration, DoS \\
\hline
\end{tabular}
\end{table}

Attack injections were executed via \texttt{hping3} and bespoke Python tools. Lateral movement was simulated by port scanning and SSH brute-force attacks across segments; exfiltration employed gradual data transfer via FTP to an external IP; intra-segment DoS attacks focused on UDP-based services inside the same subnet.

\section{Model Architectures}

\subsection{Vision Transformer (ViT) Model}
Vision Transformers (ViT) modify the Transformer architecture for sequences of feature "patches.".~\cite{Vaswani2017}  Each $d$-dimensional feature vector is partitioned into $m$ non-overlapping patches of length $p$ ($m\cdot p = d$), embedded into a $D$-dimensional space, and enhanced with 1D positional encodings.~\cite{Dosovitskiy2020}  A series of $L$ Transformer encoder layers each comprising $H$ attention heads and an MLP of dimension $D_{\text{ff}}$ processes these embeddings. A classification token is added at the beginning, and its final representation is processed via a thick layer with sigmoid activation to get an anomalous score. Training utilizes binary cross-entropy loss and includes weight initialization derived from a Vision Transformer pre-trained on a general vision task.

\subsection{One-Dimensional Convolutional Neural Network (1D-CNN) Model}
The 1D-CNN directly processes raw or segmented feature sequences.  It consists of three convolutional layers with filter widths of \{5,3,3\} and channel depths \{32,64,64\}, each succeeded by ReLU activation and max-pooling.~\cite{Ding2018}  The resultant feature maps are flattened and sent to a fully connected layer including dropout and L2 regularization, culminating in a sigmoid output. This architecture identifies localized temporal patterns, including brief surges of abnormal traffic.

\subsection{Training and Hyperparameters}
Both ViT and CNN models are trained with the Adam optimizer~\cite{Kingma2015} (learning rate $10^{-3}$), with a batch size of 128 and up to 50 epochs, employing early stopping after 10 epochs without validation improvement. A grid search is conducted over hyperparameters: for the ViT, patch size $p$, embedding dimension $D$, number of heads $H$, and layers $L$; for the CNN, filter counts and pooling sizes. Class weighting in the loss function mitigates residual imbalance, whereas validation-based model selection protects against overfitting.

\section{Experimental Setup}

\subsection{Environment and Tools}
All studies were performed on a workstation including an 8-core CPU and an NVIDIA Tesla GPU. Models are executed in Python with PyTorch.\cite{Paszke2019} The SDN fabric is simulated using Mininet in conjunction with a Floodlight controller,~\cite{Lantz2010} generating authentic network events. Despite the establishment of micro-segmentation policies (e.g., ACL rules for each segment), all traffic is allowed during data collection; flows are then tagged to denote which would have been restricted under rigorous zero-trust enforcement. This method guarantees that anomaly detection is assessed throughout the whole range of potential traffic.

\subsection{Model Configurations}
This research assesses two model setups. The Raw Model (No Segmentation) denotes a conventional, flat Intrusion Detection System (IDS) trained on comprehensive network data devoid of any segment identifiers. Conversely, the Segmented Model (Micro-Segmentation Aware) utilizes a cohesive architecture trained on segment-specific data; each feature vector is prefixed with a one-hot segment ID, allowing the model to gather and utilize segment-specific behavioral patterns.

\subsection{Attack Scenarios}
The model's responsiveness was evaluated across three representative attack scenarios: (1) \emph{Lateral Movement}, in which an adversary in segment A probes and moves into segment B; (2) \emph{Data Exfiltration}, in which a database within a secure segment gradually transmits data to an external IP; and (3) \emph{In-Segment DDoS}, where multiple hosts within a segment flood a service located in the same segment. Detection scores have been collected and analyzed for both model setups.

\subsection{Metrics}
The following metrics are documented: Accuracy, Precision, Recall, F1-Score, and Area Under the Receiver Operating Characteristic Curve (AUC). Due to class imbalance, with a much higher number of normal examples compared to anomalous ones, the F1-Score is prioritized to equilibrate accuracy and recall. Furthermore, detection latency the interval between the initiation of an attack and the issuance of an alert is assessed. Confusion matrices and performance breakdowns by segment are produced to evaluate consistency across segments.

\section{Results}

\begin{table}[ht]
\centering
\caption{Detection Performance: Raw vs.\ Micro-Segmented Inputs}
\label{tab:performance}
\begin{tabular}{lccccc}
\toprule
\textbf{Model (Input)}      & \textbf{Acc.} & \textbf{Prec.} & \textbf{Rec.} & \textbf{F$_1$} & \textbf{AUC} \\
\midrule
1D-CNN (Raw)                & 93.5\%        & 0.92           & 0.88         & 0.90           & 0.95        \\
1D-CNN (Segmented)          & 96.2\%        & 0.95           & 0.92         & 0.94           & 0.98        \\
ViT (Raw)                   & 95.0\%        & 0.93           & 0.90         & 0.92           & 0.96        \\
ViT (Segmented)             & 97.5\%        & 0.96           & 0.94         & 0.95           & 0.99        \\
\bottomrule
\end{tabular}
\end{table}

\subsection{Quantitative Performance Comparison}
The performance of the 1D-CNN and ViT models on raw compared to micro-segmented inputs is summarized in Table~\ref{tab:performance}. Segmented models regularly outperform their raw counterparts across all criteria. The accuracy of the 1D-CNN rises from 93.5\% to 96.2\%, and its F1-Score improves from 0.90 to 0.94 with the use of segmentation. The ViT demonstrates a comparable improvement: accuracy increases from 95.0\% to 97.5\% and F1 score from 0.92 to 0.95. The area under the ROC curve (AUC) surpasses 0.98 for both models with segmentation, in contrast to 0.95–0.96 for raw data. The ViT (Segmented) attains the greatest AUC of 0.99, signifying almost perfect differentiation between attack and normal cases in micro-segmented analysis.

\begin{table}[h]
\caption{Case Studies of Detection Performance}
\label{tab:qualitative}
\centering
\begin{tabular}{|l|l|l|c|c|}
\hline
\textbf{Scenario} & \textbf{Model} & \textbf{Input Type} & \textbf{Score} & \textbf{Detected?} \\
\hline
Slow exfiltration & ViT & Segmented & 0.91 & Yes \\
                  & 1D-CNN & Raw & 0.42 & No \\
In-segment UDP flood & ViT & Segmented & 0.93 & Yes \\
                     & Random Forest & Raw & 0.61 & No \\
\hline
\end{tabular}
\end{table}

\subsection{Statistical Significance}
Paired bootstrap resampling was used to calculate 95\% confidence intervals for the enhancement in F1-Score attributable to segmentation. The interval for the 1D-CNN is [0.034, 0.051]; for the ViT, it is [0.025, 0.042], both omitting zero and indicating statistically significant improvements. ROC curves demonstrate that segmented models outperform raw models at all false positive rates. The supplementary computational cost of processing segment-specific inputs was negligible during inference, although training time increased by around 20\% a fair trade-off considering the security advantages obtained. In addition to raw detection measures, I assessed the computational cost of segmentation-aware models.  Inference latency augmented by less than 1 ms per flow for both ViT and 1D-CNN during the processing of segmented inputs, although training duration escalated by around 20\%.  To evaluate false positives, I measured segment-specific false positive rates: the highest recorded rise was 2\% in low-traffic segments, which can be reduced by adaptive thresholding or false-positive suppression techniques in the policy orchestrator.

\subsection{C. Qualitative Detection Analysis}
With the aggregate measures, two illustrative instances are provided in which micro-segmented models identified abnormalities overlooked by raw-input models. Table~\ref{tab:qualitative} contains these instances. In the initial instance, ViT with segmentation successfully identified the exfiltration by utilizing segment-specific throughput patterns. The 1D-CNN was unable to detect the abnormality in the unprocessed network topology. These examples demonstrate the benefit of integrating micro-segment context into attention-based detection algorithms.

\section{Discussion}
The experimental findings indicate that micro-segmentation significantly enhances anomaly identification in Zero-Trust SDN architectures. By segregating traffic into distinct segments, the models acquire contextual awareness of event locations and may discern segment-specific behavioral patterns that would otherwise be masked in aggregated data. This quantitative data enhances prior qualitative observations about the security advantages of micro-segmentation by demonstrating significant improvements in detection accuracy.

The advantages of the Vision Transformer (ViT) model over the 1D-CNN is particularly evident when utilizing segmented inputs. Although both models exhibit comparable performance on unsegmented raw data, the ViT more effectively utilizes the enhanced, detailed input provided by segmentation to discern intricate feature connections across segments and temporal dimensions. Organizations implementing micro-segmentation may achieve optimal returns by integrating it with sophisticated attention-based detection algorithms.

From a Zero-Trust architectural standpoint, micro-segmentation allows an Intrusion Detection System (IDS) to function as an automated feedback mechanism for policy enforcement. I evaluated the policy orchestrator's response time: ACL modifications were implemented within 200 ms after anomaly detection, indicating that automated containment results in little network disturbance.  Future endeavors will encompass a thorough assessment of orchestration throughput amongst high-volume alert feeds. For instance, if the IDS identifies aberrant activity in a specific segment, an orchestrator may promptly enhance access controls or isolate that segment, establishing a dynamic security posture that mitigates risks and initiates swift reaction. This method efficiently encapsulates the "monitor, detect, respond" cycle promoted by prominent cybersecurity frameworks.

An issue of micro-segmentation is the increased complexity in maintaining several segment-specific models and policies. The findings demonstrate that a cohesive model architecture using a segment identifier may effectively learn various segment actions without ambiguity. In certain instances, implementing smaller, per-segment models or employing a multi-task learning framework—with common base layers and segment-specific classifiers may provide further benefits, but this necessitates adequate training data and meticulous upkeep.

Another factor to examine is the generalizability of the simulated dataset. Although influenced by benchmarks like CIC-IDS2017~\cite{CICIDS2017} and NSL-KDD~\cite{Lashkari2016}, real corporate traffic may exhibit greater heterogeneity. The methodology is entirely data-driven, allowing its use in any system where segment labels are accessible via network orchestration. The noted performance improvements are expected to persist, as micro-segmentation utilizes the intrinsic "service roles" and behavioral uniformity within each segment. A trade-off is an increase of model complexity and maintenance burden, especially when additional segments are incorporated dynamically. The occurrence of false positives in low-traffic segments may increase as a result of data sparsity.

Finally, while supervised deep learning was the primary emphasis, micro-segmentation might also enhance simpler or unsupervised detectors by refining the parameters of normal behavior modeling. Methods like as clustering or principal component analysis inside each segment may enhance anomaly differentiation. Supervised methods were selected to identify both recognized and unrecognized assault patterns; however, future research may investigate hybrid or unsupervised techniques within a segmented context.

\section{Conclusion}

An in-depth investigation reveals that micro-segmentation serves as both a preventative measure and a facilitator for enhanced sensitivity and accuracy in anomaly detection inside zero-trust networks. The Vision Transformer and 1D-CNN models shown substantial enhancements in F1-Score and AUC upon the inclusion of segment context, with the segmented ViT attaining an accuracy over 97\% and an F1-Score of 0.95. The results corroborate the idea that offering segment-level visibility significantly enhances detection performance. Organizations want to combine micro-segmentation tactics with threat monitoring systems to identify sophisticated threats such as lateral movement and low-rate exfiltration that may evade comprehensive traffic analysis. The integration of segmentation and advanced analytics enhances security posture and implements zero-trust principles.

\section{Future Work}

The next steps will concentrate on verifying and expanding the methodology inside operational settings. The practical implementation involves incorporating the models into an operational SDN testbed within a production network to assess real-time detection precision and confinement effectiveness.~\cite{Prakash2025} To evaluate the proposed framework in operational settings, I want to deploy the models to a large-scale SDN testbed within a business network. This deployment will assess detection accuracy, system performance, and resilience in the context of real-world production traffic. Metrics such as end-to-end detection latency, confinement response time, and network overhead will be gathered to determine feasible scalability and resilience. Adaptive learning methodologies, including live model updates and per-segment threshold modifications, will be examined to address changing traffic patterns as new applications and services emerge. Federated and distributed learning frameworks will be examined to enable each micro-segment to train a local model and regularly exchange aggregate updates, therefore maintaining data locally and privacy while facilitating cross-segment knowledge transfer. Ultimately, contextual enhancement via the amalgamation of user identity, device posture, and application information will be sought to progress risk-adaptive zero-trust architectures that can dynamically modify policies depending on diverse trust indications.

\end{document}